\documentclass[letter,traditabstract]{aa}
\newcommand{\GILDAS}{\texttt{GILDAS}}
\usepackage{natbib}
\usepackage{amsmath}
\usepackage{graphicx}
\usepackage{txfonts}
\usepackage{subfig}
\usepackage{multirow} 
\usepackage{threeparttable}
\usepackage{color}
\newcommand{\ie}{\emph{i.e.}}
\newcommand{\eg}{e.g.}
\newcommand{\emm}[1]{\ensuremath{#1}}   % Ensures math mode.
\newcommand{\emr}[1]{\emm{\mathrm{#1}}} % Uses math roman fonts.

\newcommand{\Tex}{\emm{T_\emr{ex}}}

\newcommand{\nH}{\emm{n_\emr{H}}}
\newcommand{\nHH}{n\emr{(H_2)}}
\newcommand{\OI}{O\,{\sc i}}
\newcommand{\Cp}{\emr{C^{+}}}                  % C+
\newcommand{\CFp}{\emr{CF^{+}}}                    % CF+
\newcommand{\DCOp}{\emr{DCO^{+}}}                  % DCO+
\newcommand{\hh}{\text{H}_2}                       % H2
\renewcommand{\deg}{\emm{^\circ}}
\newcommand{\pccm}{~\rm{cm}^{-3}}
\newcommand{\pscm}{~\rm{cm}^{-2}}
\newcommand{\ps}{~\rm{s}^{-1}}
\newcommand{\kms}{~\rm{km}~\mathrm{s}^{-1}}
\newcommand{\Kkms}{~\rm{K\,km\,s}^{-1}}
\newcommand{\Tsys}{\emm{T_\emr{sys}}}
\newcommand{\Tas}{\emm{T_\emr{A}^*}}
\newcommand{\Tmb}{\emm{T_\emr{mb}}}
\newcommand{\Beff}{\emm{B_\emr{eff}}}
\newcommand{\Feff}{\emm{F_\emr{eff}}}

\begin{document}
%
%
%\title{$\CFp$ detection in the Horsehead: \\A tracer of $\Cp$, the
%  outermost layers in PDRs}
%\title{$\CFp$ in the Horsehead: A tracer of $\Cp$ and a
%  direct measurement of the Fluorine abundance in PDRs}
\title{The IRAM-30m line survey of the Horsehead PDR:\\ I. $\CFp$ as a
  tracer of $\Cp$ and a measure of the Fluorine abundance}
%\title{Horsehead WHISPER: $\CFp$ as a tracer of $\Cp$ \\and a measure of
%  the Fluorine abundance}

\authorrunning{V. Guzm\'an, J. Pety, P. Gratier et al.}
\titlerunning{$\CFp$ as a tracer of $\Cp$ and a measure of the
  Fluorine abundance}

% \subtitle{}

   \author{V. Guzm\'{a}n\inst{1} \and J. Pety\inst{1,2} \and
     P. Gratier\inst{1} \and J.R. Goicoechea\inst{3} \and
     M. Gerin\inst{2} \and E. Roueff\inst{4} \and D. Teyssier\inst{5} }

          \institute{
            IRAM, 300 rue de la Piscine, 38406 Saint Martin d'H\`eres, France\\
            \email{[guzman;pety;gratier]@iram.fr}
            \and
            LERMA - LRA, UMR 8112, Observatoire de Paris and Ecole 
            normale Sup\'{e}rieure, 24 rue Lhomond, 75231 Paris, France. \\
            \email{maryvonne.gerin@lra.ens.fr}       
            \and
            Centro de Astrobiolog\'{i}a. 
            CSIC-INTA. Carretera de Ajalvir, Km 4. Torrej\'{o}n de Ardoz, 
            28850 Madrid, Spain. \\
            \email{jr.goicoechea@cab.inta-csic.es}
            \and
            LUTH UMR 8102, CNRS and Observatoire de Paris, Place J. Janssen, 
            92195 Meudon Cedex, France.\\
            \email{evelyne.roueff@obspm.fr}
            \and
            European Space Astronomy Centre, ESA, PO Box 78, 28691
            Villanueva de la Cañada, Madrid, Spain\\ 
            \email{dteyssier@sciops.esa.int}
          }

%%    \date{Received September 15, 1996; accepted March 16, 1997}

\abstract{$\Cp$ is a key species in the interstellar medium but its
  158~$\mu$m fine structure line cannot be observed from ground-based
  telescopes. Current models of fluorine chemistry predict that $\CFp$
  is the second most important fluorine reservoir, in regions where
  $\Cp$ is abundant. We detected the $J=1-0$ and $J=2-1$ rotational
  lines of $\CFp$ with high signal-to-noise ratio towards the PDR and
  dense core positions in the Horsehead. Using a rotational diagram
  analysis, we derive a column density of N$(\CFp) = (1.5-2.0) \times
  10^{12} \pscm$. Because of the simple fluorine chemistry, the \CFp{}
  column density is proportional to the fluorine abundance. We thus
  infer the fluorine gas-phase abundance to be $\mbox{F/H} =
  (0.6-1.5)\times10^{-8}$. Photochemical models indicate that $\CFp$
  is found in the layers where $\Cp$ is abundant. The emission arises
  in the UV illuminated skin of the nebula, tracing the outermost
  cloud layers.  Indeed, $\CFp$ and $\Cp$ are the only species
  observed to date in the Horsehead with a double peaked line profile
  caused by kinematics. We therefore propose that $\CFp$, which is
  detectable from the ground, can be used as a proxy of the $\Cp$
  layers.}

     %{}
     % aims heading (mandatory)
     %{}
     % methods heading (mandatory)
     %{}
     % results heading (mandatory) 
     %{}
     % conclusions heading (optional), leave it empty if necessary 
     %{}
%
   \maketitle
%
%________________________________________________________________

\newcommand{\TabObs}{%
  \begin{table*}
    \begin{center}
    {\small \caption{Observation parameters for the maps. Their
        projection center is $\alpha_{2000} = 05^h40^m54.27^s$,
        $\delta_{2000} = -02\deg 28' 00''$.}} 
    \label{tab:obs:maps}
      %{\tiny
        \begin{tabular}{lrlcccccccr}
          \hline \hline
          Line & Frequency & Instrument & $F_{\textrm{eff}}$ &
          $B_{\textrm{eff}}$ & Beam & PA & Int. Time$^a$ & \Tsys{} & Noise
          & Obs.date\\  
          & GHz & & & & arcsec & $^{\deg}$ & hours & K (\Tas{}) & K
          (\Tmb{}) & \\ 
          \hline
          HCO $1_{0,1}\,3/2, 2 - 0_{0,0}\,1/2,1$ & 86.670760 & 
          PdBI/C\&D & 0.95 & 0.78 & $6.7 \times 4.4$ & 16 & 6.5 & 150 &
          0.09 & 2006-2007 \\ 
          \CFp{} $1-0$ & 102.587533 & 30m/EMIR & 0.94 & 0.79 & 25.4 &
          0 & 2.5 & 88 & 0.03$^{b}$ & Jan. 2012 \\   
          \CFp{} $2-1$ & 205.170520 & 30m/EMIR & 0.94 & 0.64 & 11.4 &
          0 & 3.4 & 220 & 0.18$^{c}$ & Jan. 2012 \\ 
          \DCOp{} $3-2$  & 216.112582 & 30m/HERA & 0.90 & 0.52 & 11.4
          & 0 & 1.5 & 230 & 0.10 & Mar. 2006 \\   
          \hline
        \end{tabular}%}
    \end{center}
    $^{a}$ On-source integration time. The noise values are for a
    spectral resolution of $^{b}$~0.114$\kms$ and $^{c}$~0.057$\kms$.
\end{table*}
}

\newcommand{\TabGaussianFit}{%
\begin{table}[b!]
  \begin{center}
  \caption{Gaussian fit results.} 
  \label{tab:obs:fit}
  \begin{tabular}{lcccc}
    \hline 
    \hline
    Line & Line area & Velocity & Width & $T_{\textrm{peak}}$ \\
    & $\Kkms$ & $\kms$ & $\kms$ & K \\
    \hline
    & \multicolumn{4}{c}{PDR}\\
    % line                                Area             Velocity             Width       Tpeak   
    \multirow{2}{*}{$\CFp J=1-0$} & 0.10$\pm$0.01 & 10.36$\pm$0.02 & 0.65$\pm$0.04 & 0.15\\
                 & 0.05$\pm$0.01 & 11.38$\pm$0.03 & 0.66$\pm$0.09 & 0.07\\
    \multirow{2}{*}{$\CFp J=2-1$} & 0.25$\pm$0.02 & 10.62$\pm$0.02 & 0.67$\pm$0.06 & 0.36\\
                 & 0.04$\pm$0.02 & 11.39$\pm$0.06 & 0.38$\pm$0.17 & 0.09\\
    \hline                                           
    % CORE
    & \multicolumn{4}{c}{CORE}\\
    \multirow{2}{*}{$\CFp J=1-0$} & 0.03$\pm$0.01 & 10.47$\pm$0.04 & 0.60$\pm$0.08 & 0.05\\ 
                 & 0.01$\pm$0.01 & 11.32$\pm$0.06 & 0.45$\pm$0.20 & 0.03\\
    $\CFp J=2-1$ & $<0.05^a$ & 10.7$^a$ & 0.5$^a$ & 0.09$^a$\\
    \hline
  \end{tabular}
  \end{center}
  $^a$ Upper limit for a fixed velocity, linewidth and a
  $T_{\textrm{peak}} = 2 \sigma_{\textrm{RMS}}$
\end{table}
}

\newcommand{\FigLines}{%
  \begin{figure}[t!]
    \centering
    \includegraphics[scale=0.48]{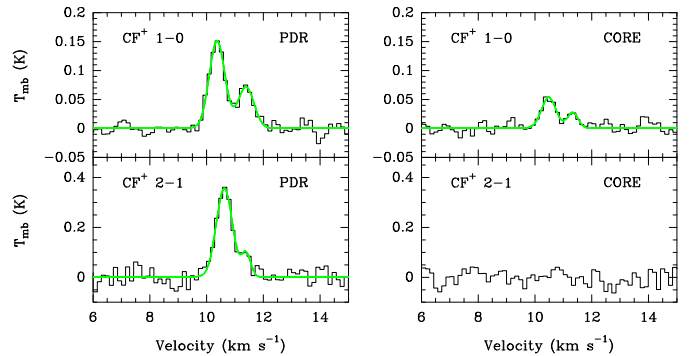}
    \caption{\small{Deep integrations towards the PDR and core positions. The
      green lines are double gaussian fits.} }
    \label{fig:lines}
  \end{figure}
}

\newcommand{\FigMaps}{%
  \begin{figure}[t!]
    \centering
      \includegraphics[scale=0.42]{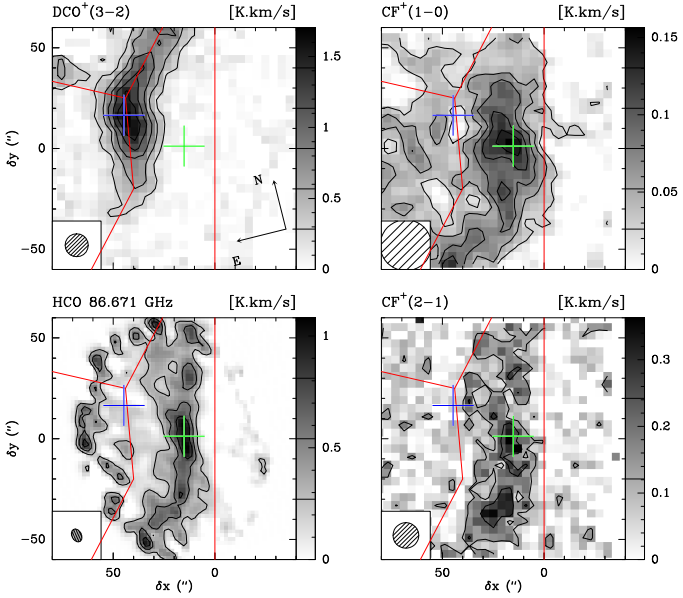}
      \caption{\small{Integrated intensity maps of the Horsehead edge.
          Maps were rotated by 14$\deg$ counter-clockwise around the
          projection center, located at ($\delta x,\delta y$) = (20",
          0′"), to bring the exciting star direction in the horizontal
          direction and the horizontal zero was set at the PDR edge,
          delineated by the red vertical line. The crosses show the
          positions of the PDR (green) and the dense core (blue). The
          spatial resolution is plotted in the bottom left
          corner. Values of the contour levels are shown on the color
          look-up table of each image (first contour at $2\sigma$ and
          $2.5\sigma$ for $\CFp 1-0$ and $2-1$ respectively). The
          emission of all lines is integrated between 10.1 and 11.1
          $\kms$.}}
      \label{fig:maps}
  \end{figure}
}

\newcommand{\FigModel}{%
  \begin{figure}[t!]
    \centering
    \includegraphics[scale=0.75]{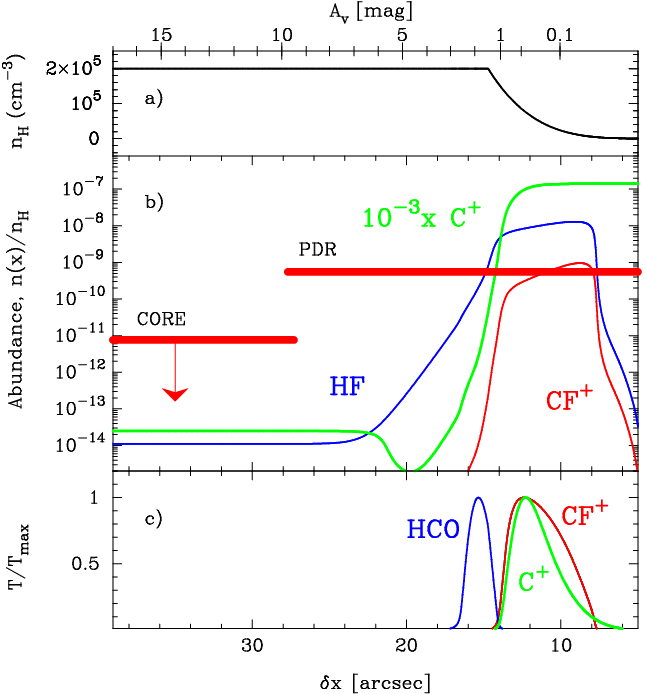} 
    \caption{\small{Photochemical model of the Horsehead PDR. $A_V$
        increases from right to left and the PDR edge, delineated by
        the red vertical line in Fig.~2, corresponds to $A_V=0$. a)
        Horsehead density profile $\nH = n$(H) + $2\nHH$. b) Predicted
        abundance of $\CFp$ in red, HF in blue and $\Cp$ in green. The
        red horizontal bars show the measured $\CFp$ abundances, and
        their length represents the beam size. c) Predicted intensity
        profiles.}}
    \label{fig:model}
  \end{figure}
}

%________________________________________________________________________

\section{Introduction}

$\Cp$ is a key species in the interstellar medium. First, it is an
important tracer of the neutral gas where CO has not been able to form
yet \citep{langer10}. Second, it is the dominant gas phase reservoir
of carbon in the diffuse interstellar medium and its fine structure
transition ($^2P_{3/2}$ $-$ $^2P_{1/2}~157.8~\mu$m, 1.9~THz) is the
major cooling mechanism of the diffuse gas. Unfortunately, its
rest-frame emission cannot be observed from ground-based
telescopes. With Herschel and SOFIA it is now possible to observe this
line from space and from the stratosphere respectively, but with
limited spatial resolution ($12"$ and $15"$, respectively). It is
therefore of great interest to find tracers of $\Cp$ that can be
observed from the ground at much higher spatial resolution, for
example with the Atacama Large Millimeter Array (ALMA) or the Northern
Extended Millimeter Array (NOEMA).

The chemistry of fluorine was studied by \cite{neufeld05}. Although
there are still uncertainties in some reaction rate coefficients
(\eg{}~$\CFp$ photodissociation), the current models predict that
$\CFp$ is present in regions where $\Cp$ and HF are abundant, as it is
produced by reactions between these species and destroyed mainly by
electrons. In these regions $\CFp$ is the second most important
fluorine reservoir, after HF. The ground rotational transition of HF,
which lies at THz frequencies, was first detected with the
HIFI/Herschel in the diffuse interstellar medium through absorption
measurements \citep{neufeld10} and in emission in the Orion Bar
\citep{vandertak12}. \cite{neufeld10} find a lower limit for the HF
abundance of $6\times10^{-9}$ relative to hydrogen nuclei, providing
support to the theoretical prediction that HF is the dominant fluorine
reservoir under a wide variety of interstellar conditions. Unlike HF
and $\Cp$, $\CFp$ rotational lines can be observed from the
ground. However, up to now there has been only one detection of $\CFp$
towards the Orion Bar \citep{neufeld06}.

In this letter we report the detection of $\CFp$ at the HCO and
\DCOp{} peak emission, corresponding respectively to the PDR and dense
core environments in the Horsehead nebula \citep{pety07,gerin09}. We
then infer the gas phase fluorine abundance.

\TabObs{} %

\vspace{-0.2cm}

\section{Observations and data reduction}
\label{sec:obs}

Fig.~\ref{fig:lines} displays deep integrations of the $J=1-0$ and
$J=2-1$ low-energy rotational lines of $\CFp$ with the IRAM-30m
telescope centered at the PDR and the dense core, located
  respectively at ($\delta$RA, $\delta$DEC) = $(-5'',0'')$ and
  $(20'',22'')$ with respect to the projection center, with 49~kHz
spectral resolution at both frequencies. These observations were
  obtained as part of the Horsehead WHISPER project (Wideband
  High-resolution Iram-30m Surveys at two Positions with Emir
  Receivers). A presentation of the whole survey and the data
reduction process will be given in Pety et al. 2012, in prep.

The $\CFp$ $J=1-0$ and $J=2-1$ maps displayed in Fig.~\ref{fig:maps}
were observed simultaneously during 7 hours of good winter weather
(2mm of precipitable water vapor) using the EMIR sideband-separation
receivers at the IRAM-30m. We used the position-switching, on-the-fly
observing mode. The off-position offsets were ($\delta$RA,
$\delta$Dec) = $(-100'',0'')$, i.e. into the HII region ionized by
$\sigma$Ori and free of molecular emission. We observed along and
perpendicular to the direction of the exciting star in zigzags,
covering an area of $100''\times100''$. A description of the HCO and
\DCOp{} observations and data reductions, which are also displayed in
Fig.~\ref{fig:maps}, can be found in \cite{gerin09} and \cite{pety07}.
Table 1 summarizes the observation parameters for all these maps.

The maps were processed with the \GILDAS{}\footnote{See
  \texttt{http://www.iram.fr/IRAMFR/GILDAS} for more information about
  the \GILDAS{} softwares.} softwares~\citep{pety05}.  The IRAM-30m
data were first calibrated to the \Tas{} scale using the chopper-wheel
method~\citep{penzias73}, and finally converted to main-beam
temperatures (\Tmb{}) using the forward and main-beam efficiencies
(\Feff{} \& \Beff{}) displayed in Table~1. The resulting amplitude
accuracy is $\sim 10\%$. The resulting spectra were then
baseline-corrected and gridded through convolution with a Gaussian to
obtain the maps.

\vspace{-0.2cm}

\section{Results}
\label{sec:results}

\subsection{Line profiles}

Two velocity peaks for the $J=1-0$ line are clearly seen at both
positions in Fig.~\ref{fig:lines}. The second velocity peak is
marginal for the $J=2-1$ line. Table~\ref{tab:obs:fit} presents the
results of dual Gaussian fits. The centroid velocity of each peak is
significantly shifted between the two $\CFp$ transitions. We have
carefully checked that neither peak is a residual line incompletely
rejected from the image side band. This doubled-peak behavior
is unexpected because all species without an hyperfine structure
detected previously at millimeter wavelengths in the 
Horsehead present a simple velocity profile centered close to
$10.7~\kms$. The only other species detected to date with a clear
double peak emission profile is $\Cp$ towards the illuminated edge of
the cloud (HIFI/Herschel, Teyssier et al. 2012 in prep).

The most obvious explanation would be that the higher velocity peak
corresponds to another line from another species. However, there are
no other lines in the public line catalogs \citep{picket98,muller01}
near this frequency besides $\CFp$. Another possible explanation would
be that the two peaks correspond to different hyperfine components
which are caused by the fluorine nuclei. To our knowledge, there are
no hyperfine structure studies on $\CFp$. However, as the molecule is
isoelectronic with CO, one can try to rely on $^{13}$CO spectroscopy
since the nuclear spin of $^{13}$C is 1/2, as for the fluorine
nucleus. The magnetic dipolar coupling constant scales approximately
with the rotational constant and the magnetic moment of the nucleus
for $^1 \Sigma$ electronic ground states \citep{reid74}. This allows
us to derive a coupling constant of approximately 110~kHz, and an
hyperfine splitting of 165~kHz (110~kHz) for the $J=1-0$ ($J=2-1$)
transitions, well below the observations. In addition, the respective
intensities do not follow the theoretical predictions. Therefore, this
possibility is unlikely. The profile is not caused by self-absorption
because the $\CFp$ opacities are low ($\tau \la 1$). We thus attribute
the complex line profiles to kinematics in the $\CFp$ emitting layers.

\FigLines{}
\FigMaps{}
\TabGaussianFit{}%

%\vspace{-0.2cm}

\subsection{$\CFp$ spatial distribution}

Fig.~\ref{fig:maps} presents the $\CFp$, HCO and \DCOp{} integrated
line intensity maps. The $\CFp$ emission is concentrated towards the
edge of the Horsehead, delineating the western edge of the \DCOp{}
emission. A more extended and fainter emission is detected in the
$\CFp~J=1-0$ map but not in the $J=2-1$ map, which has a lower
signal-to-noise ratio. The intensity peak of the $\CFp J=1-0$
line coincides with the intensity peak of the HCO emission
(shown by the green cross), which traces the far UV illuminated matter
\citep{gerin09}. The $J=2-1$ transition also peaks near the HCO
emission peak. %, but it presents two more maxima towards the south. On
%the other side, the beam size is considerably larger for the $\CFp$
%lines than for the HCO PdBI map.

We have checked that beam pick-up contamination from the PDR is
negligible at the core position ($<7\%$), even with the large beam
size ($\sim25"$) of the 30m at 102~GHz. The emission then arises in
the line of sight towards the core but not necessarily in the cold gas
associated with the core. Indeed, we expect the $\CFp$ emission to
arise in the outer layers of the nebula, delineating the edge as shown
by the maps. This emission is likely to arise in the warmer and more
diffuse material of the skin layers towards the core line of sight.
Furthermore, there is a minimum of the emission in the $\CFp 1-0$ map
towards the dense core position. This confirms that the $\CFp$
emission is associated to the diffuse envelope. Molecular emission
from the lower density cloud surface was already mentioned by
\cite{goicoechea06} and \cite{gerin09} to explain the CS and HCO
emission, respectively.

\subsection{Column densities and abundances}

The $\CFp$ column density is estimated assuming that the emission is
optically thin and that the emission fills the 30m beam. We infer an
excitation temperature of 10~K, based on a rotational diagram built
with the integrated line intensities of the two detected transitions.
Assuming $\Tex=10$~K for all rotational levels, the beam averaged
column density is $\simeq (1.5-2.0)\times10^{12} \pscm$ in the PDR.
This value is similar to the column density found in the Orion Bar by
\cite{neufeld06}. In the next section, we will show that the $\CFp$
emission arises at the illuminated edge of the
nebula. \cite{goicoechea09a} found that the [\OI]63~$\mu$m fine
structure line, which also arises at the edge of the nebula, was best
reproduced with a gas density of $\nH \simeq 10^4 \pccm$. Thus,
assuming this density and a cloud depth $l\sim 0.1$~pc
\citep[][]{habart05}, the $\CFp$ column density translates into an
abundance of $\simeq(4.9-6.5)\times10^{-10}$ with respect to H
nuclei. Taking the same excitation temperature of $10$~K, we computed
a column density towards the core of $\simeq 4.4\times10^{11}
\pscm$. We consider this as an upper limit for the model in
Sect.~\ref{sec:chem} because $\CFp$ is found in the surface layer
which is not taken into account by our unidimensional photochemical
model.

\subsection{$\CFp$ chemistry}
\label{sec:chem}

\begin{figure}[b!]
  \centering
  \includegraphics[scale=0.45]{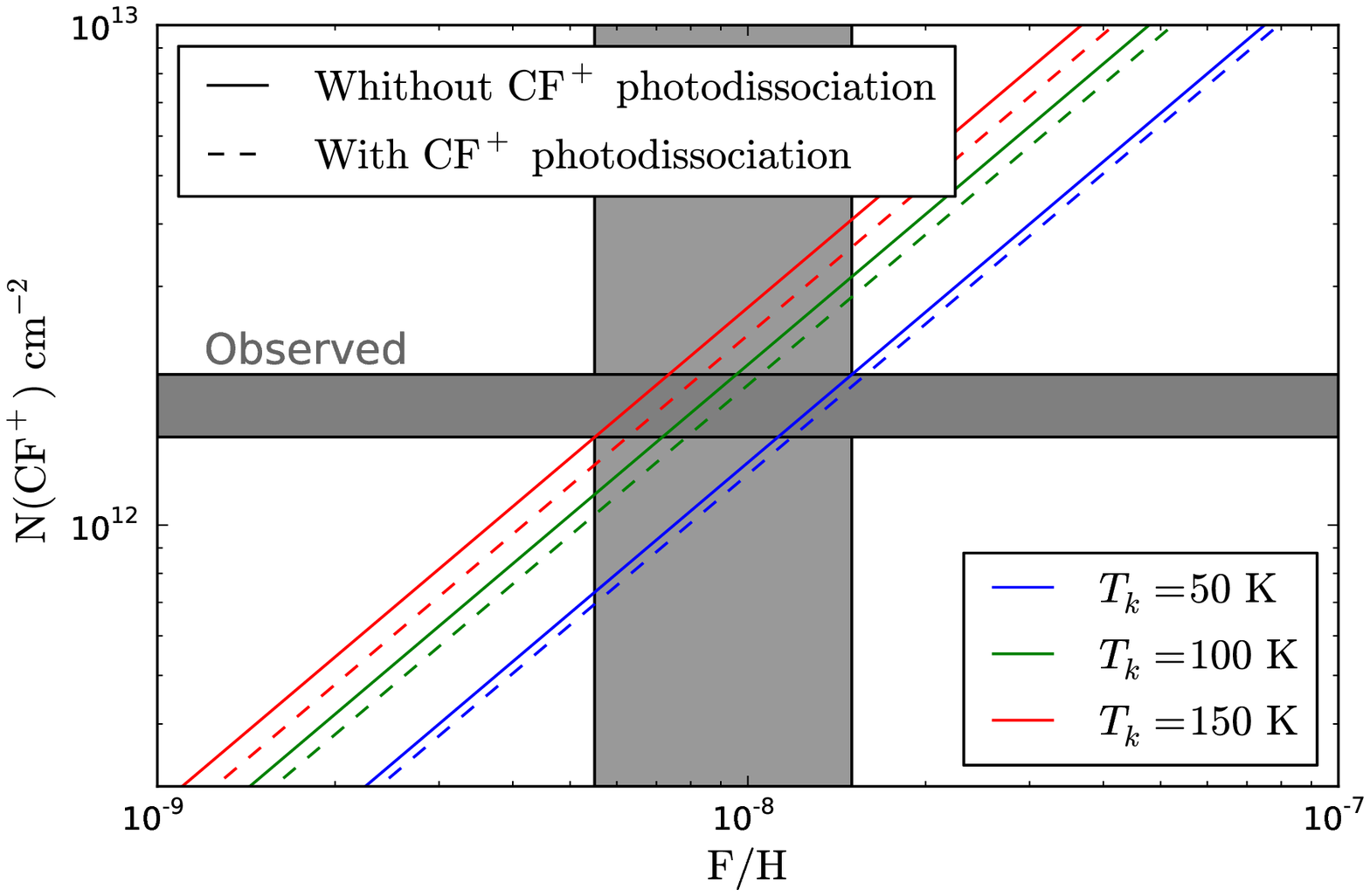}
  \caption{\small{Relation between the \CFp{} column density and
      F/H.}}
  \label{fig:Fabund}
\end{figure}

$\CFp$ is formed through the following chemical path:
\begin{align*}
\textrm{F} + \hh &\xrightarrow[]{} \textrm{HF} + \textrm{H} \\
\textrm{HF} + \Cp &\xrightarrow[]{k_1} \CFp +  \textrm{H} 
\end{align*}
and destroyed by dissociative recombination with electrons
\citep{neufeld06} and by far UV photons:
\begin{align*}
\CFp + \textrm{e}^{-} &\xrightarrow[]{k_2} \textrm{C} + \textrm{F}\\ 
\CFp + h\nu &\xrightarrow[]{k_{\textrm{pd}}} \Cp + \textrm{F} 
\end{align*}
The reactions rates are $k_1 =
7.2\times10^{-9}(T/300)^{-0.15}$cm$^3$~s$^{-1}$ \citep{neufeld05} and
$k_2 = 5.3\times10^{-8}(T/300)^{-0.8}$ cm$^3$~s$^{-1}$
\citep{novotny05}. The \CFp{} photodissociation rate
$k_{\textrm{pd}}$ is not known. Nevertheless, assuming a rate
of $10^{-9}~\exp(-2.5~A_V) \ps$, we estimate that this contribution is
negligible compared to the dissociative recombination in low far UV
field PDRs like the Horsehead. However, it might not be negligible in
regions with high radiation fields ($\chi \simeq 10^4-10^5$), like the
Orion Bar. In the following we thus assume that $k_{\textrm{pd}}=0$.
Because 1) the fluorine chemistry is simple, 2) the electron abundance
is given by the ionization of carbon $n(\mathrm{e}-) \sim n(\Cp)$ and 3) HF is
the major reservoir of fluorine $n(\textrm{HF}) \sim n(\textrm{F})$,
it can be shown that the \CFp{} column density is proportional to the
fluorine gas phase abundance ([F] = F/H), \ie{}
\begin{equation*}
\mathrm{N}(\CFp) \simeq \frac{k_1}{k_2}
       [\mathrm{F}]~\nH~l \hspace{0.5cm} [\pscm]
\end{equation*}
 From our \CFp{} observations we find F/H $\simeq
 (0.6-1.5)\times10^{-8}$ in the Horsehead PDR (see
 Fig.~\ref{fig:Fabund}), in good agreement with the solar value
 \citep[$2.6\times10^{-8}$;][]{asplund09} and the one found in the
 diffuse atomic gas
 \citep[$1.8\times10^{-8}$;][]{snow07}. \cite{sonnentrucker10} also
 derived F/H $\simeq (0.5-0.8)\times10^{-8}$ in diffuse molecular
 clouds detected in absorption with the HIFI/Herschel.
 
In order to understand the $\CFp$ abundance profile as a function of
depth, we used a one-dimensional, steady-state photochemical model
\citep[Meudon PDR code,][]{lebourlot12,lepetit06}. The used version of
the Meudon PDR model includes the Langmuir Hinshelwood and Eley-Rideal
mechanisms to describe the formation of $\hh$ on grain surfaces. The
physical conditions in the Horsehead have already been constrained by
our previous observational studies and we keep the same assumptions
for the steep density gradient (displayed in the upper panel of
Fig.~\ref{fig:model}), radiation field ($\chi = 60$ times the
\cite{draine78} mean interstellar radiation field), elemental
gas-phase abundances \citep[see Table 6 in][]{goicoechea09b} and
cosmic ray primary ionization rate ($\zeta= 5\times10^{-17}~\ps$ per
$\hh$ molecule). We used the Ohio State University (osu) pure
gas-phase chemical network, and included fluorine adsorption and
desorption on grains.

The predicted $\CFp$, HF and $\Cp$ abundance profiles are shown in
Fig.~\ref{fig:model}~(b). HF and $\CFp$ abundances decreases rapidly
for $A_V>1$.  The model is in good agreement with the observed $\CFp$
abundance in the PDR, shown by the horizontal bars. The model predicts
that there is a significant overlap between $\CFp$ and
$\Cp$. Moreover, the abundance ratio between these two species remains
quite constant along the illuminated side of the cloud, \ie{} $A_V<4$,
as shown in Fig.~\ref{fig:model}~(c). The $\CFp$ emission arises in
the outermost layers of the cloud ($A_V \sim 0.5$), which are directly
exposed to the far UV field and where the gas is less dense. The
predicted spatial distribution of the $\CFp$ emission is shown in
Fig.~\ref{fig:model}~(d). We expect a narrow filament ($\sim 5''$)
shifted in the illuminated part of the PDR with respect to the HCO
emission, which has already shown to trace the far UV illuminated
molecular gas \citep{gerin09}.

\vspace{-0.2cm}

\section{Conclusions}

We have detected the $J=1-0$ and $J=2-1$ rotational lines of $\CFp$
with high signal-to-noise ratio towards the PDR and core positions in
the Horsehead. We have also mapped the region, and find that the
emission arises mostly at the illuminated edge of the nebula (PDR),
but it is also detected towards the dense core arising from its lower
density skin. \CFp{} is unique as its column density is proportional
to the elemental abundance of fluorine. In the Horsehead PDR we find
$N \simeq (1.5-2.0) \times 10^{12} \pscm$ and infer
F/H~$\simeq(0.6-1.5)\times10^{-8}$. Our model of the fluorine
chemistry predicts that $\CFp$ accounts for 4-8\% of all fluorine.
$\CFp$ is found in the layers where $\Cp$ is abundant as it is formed
by reactions between HF and $\Cp$. In these regions the ionization
fraction is high \citep[see ][]{goicoechea09b} and $\CFp$ destruction
is dominated by dissociative recombination with electrons. The $\CFp$
emission has two velocity components. The possibility that we are
resolving the hyperfine structure is unlikely but corresponding
theoretical or experimental study would allow to derive the velocity
structure unambiguously. Although the $\CFp$ line profile is not
exactly the same as the $\Cp$ line profile, they are the only species
in the Horsehead with a double-peaked profile of kinematic origin
measured to date. The complex line profile of both \CFp{} and $\Cp$
therefore confirms that they trace the gas directly exposed to the far
UV radiation, which shows a completely different kinematics than the
following layers traced by other species, like HCO. We therefore
propose that $\CFp$ can be used as a proxy of the $\Cp$ layers which
can be observed from the ground. We will check this by comparing with
a HIFI/Herschel map of the $\Cp$ emission in the Horsehead.

\FigModel{}

\begin{acknowledgements}
  We thank the editor and anonymous referee for their useful comments
  that improved the letter.  VG thanks support from the Chilean
  Government through the Becas Chile scholarship program. This work
  was also funded by grant ANR-09-BLAN-0231-01 from the French {\it
    Agence Nationale de la Recherche} as part of the SCHISM
  project. JRG thanks the Spanish MICINN for funding support through
  grants AYA2009-07304 and CSD2009-00038. JRG is supported by a
  Ram\'on y Cajal research contract from the Spanish MICINN and
  co-financed by the European Social Fund.
\end{acknowledgements}

\vspace{-0.2cm}

\bibliographystyle{aa}
\vspace{-0.2cm}
\bibliography{draft-cfp}

\end{document}